\documentclass[aps,prl,twocolumn,superscriptaddress,showpacs]{revtex4}

\usepackage{graphicx}
\usepackage{amsmath,amssymb}
\usepackage{color}

\begin{document}

\title{Nonergodic dynamics of force-free granular gases}

\author{Anna Bodrova}
\affiliation{Institute of Physics and Astronomy, University of Potsdam,
D-14476 Potsdam-Golm, Germany}
\affiliation{Faculty of Physics, M. V. Lomonosov Moscow State University,
Moscow 119991, Russia}
\author{Aleksei V. Chechkin}
\affiliation{Institute of Physics and Astronomy, University of Potsdam,
D-14476 Potsdam-Golm, Germany}
\affiliation{Akhiezer Institute for Theoretical Physics, Kharkov Institute
of Physics and Technology, Kharkov 61108, Ukraine}
\author{Andrey G. Cherstvy}
\affiliation{Institute of Physics and Astronomy, University of Potsdam,
D-14476 Potsdam-Golm, Germany}
\author{Ralf Metzler}
\affiliation{Institute of Physics and Astronomy, University of Potsdam,
D-14476 Potsdam-Golm, Germany}
\affiliation{Department of Physics, Tampere University of Technology,
33101 Tampere, Finland}

\begin{abstract}
We study analytically and by event-driven molecular dynamics simulations the
nonergodic and aging properties of force-free cooling granular gases with both
constant and velocity-dependent (viscoelastic) restitution coefficient
$\varepsilon$ for particle pair collisions. We compare the granular gas dynamics
with an effective single particle stochastic model based on an underdamped
Langevin equation with time dependent diffusivity. We find that both models share
the same behavior of the ensemble mean squared displacement (MSD) and the velocity
correlations in the small dissipation limit. However, we reveal that the time
averaged MSD of granular gas particles significantly differs from this effective
model due to ballistic correlations for systems with constant $\varepsilon$.
For velocity-dependent $\varepsilon$ these corrections become weaker at
longer times. Qualitatively the reported
non-ergodic behavior is generic for granular gases with any realistic dependence
of $\varepsilon$ on the impact velocity.
\end{abstract}


\maketitle

Granular materials such as stones, sand, different types of powders, or their
mixtures are ubiquitous in Nature and technology, for instance, in the cosmetic,
food, and building industries \cite{DryGranMed}. Rarefied granular systems, in
which the distance between particles exceeds their size, are called granular
gases \cite{book}. On Earth granular gases may be realized by placing granular
matter into a container with vibrating \cite{wildman} or rotating \cite{rotdriv}
walls, applying electrostatic \cite{electrodriven} or magnetic \cite{magndriven}
forces, etc. Granular gases are common in Space, occurring in protoplanetary
discs, interstellar clouds, and planetary rings (e.g., Saturn's) \cite{PlanRings}.
In physics granular gases are of fundamental importance: they are direct
generalizations of ideal gases when the restriction of ideal collisions is
dropped \cite{book}.

Ergodicity is a fundamental concept of statistical mechanics. Starting with
Boltzmann, the ergodic hypothesis states that long time averages of physical
observables are identical to their ensemble averages \cite{ergo}. In the wake of
modern
microscopic techniques such as single particle tracking \cite{brauchle}, in which
individual trajectories of single molecules or submicron tracers are routinely
measured, knowledge of the ergodic properties of the system is again pressing:
while time averages are measured in single particle assays or massive simulations,
generally ensemble averages are more accessible theoretically. How measured time
averages can be interpreted in terms of ensemble approaches and diffusion models
is thus imminent \cite{m1,mreview}.

Here we demonstrate in quantitative detail how ergodicity is violated even in
simple mechanical systems
such as force-free granular gases. We analytically derive the time and ensemble
averaged mean squared displacements (MSDs) and show that for both constant and
viscoelastic restitution coefficients the time average of the MSD is
fundamentally different from the corresponding ensemble MSD. Moreover, the
amplitude of the time averaged MSD is shown to be a decaying function of the length
of the measured trajectory (aging). Comparison to the effective single particle
underdamped scaled Brownian motion (SBM) demonstrates that this behavior is due to
the non-stationarity invoked by the time dependence of the granular temperature.
Ballistic correlations of the granular gas are shown to be relevant beyond this
effective SBM description. Our results for generic granular gases
are relevant both from a fundamental statistical mechanical point of view and
for the practical analysis of time series of granular gas particles from
observations (in particular, of granular gases in Space) and simulations.

Granular gas particles collide inelastically and a fraction of their kinetic
energy is transformed into heat stored in internal degrees of freedom. The
dissipative nature of granular gases gives rise to many interesting physical
properties \cite{book}. In absence of external forces the gas gradually cools down.
During the first stage of its evolution, the granular gas is in the homogeneous
cooling state characterized by uniform density and absence of macroscopic fluxes
\cite{book}, realized, e.g., in microgravity environments \cite{expgg}. Eventually
instabilities occur and vertexes develop \cite{book}. Here we focus on spatially
uniform systems.

\begin{figure}
\includegraphics[width=7.2cm]{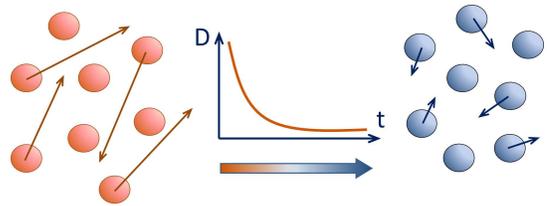}
\caption{Collisions in a free granular gas with a below-unity restitution
coefficient lead to a cooling of the gas. Along with the reduced kinetic
energy of the gas particles, the diffusion coefficient in a free cooling
granular gas decreases with time.}
\label{grangas}
\end{figure}

Collisions are quantified by the restitution coefficient $\varepsilon=\left|v'_{
12}/v_{12}\right|$, the ratio of the relative speed of two granular particles after
and before a collision event \cite{REMM}; $\varepsilon=1$ denotes perfectly elastic
collisions, while $\varepsilon=0$ reflects the perfectly inelastic case \cite{RRR}.
For $0<\varepsilon<1$ the granular temperature $T(t)=m\langle\mathbf{v}^2\rangle/2$
given by the mean kinetic energy of particles with mass $m$ continuously decreases
according to Haff's law for granular gases \cite{Haff},
\begin{equation}
T(t)=T_0/\left(1+t/\tau_0\right)^2.
\label{haff}
\end{equation}
Here $\tau_0^{-1}=\frac16\left(1-\varepsilon^2\right)\tau_{c}^{-1}(0)$ is the
inverse characteristic time of the granular temperature decay, involving the
initial value of the inverse mean collision time $\tau_c^{-1}(t)\propto\sqrt{
T(t)/m}$. Weak dissipation ($\varepsilon\simeq1$) thus implies $\tau_0 \gg\tau_c$.
Due to the temperature decrease the self-diffusion coefficient $D(t)$ of the gas
is time dependent \cite{book,seldiffusionbril,seldiffusionbrey,anna_gran},
\begin{equation}
D(t)=T(t)\tau_v(t)/m=D_0/(1+t/\tau_0)
\end{equation}
with $\tau_v(t)=3\tau_c(t)/2$, $D_0=T_0\tau_v(0)/m$ (Fig.~\ref{grangas}). For
$\varepsilon=1$ we recover normal diffusion with constant diffusivity.

Most studies of granular gases assume that $\varepsilon$ is constant. Different
approaches consider the relative collision speed dependence of $\varepsilon$ as
$\varepsilon(v_{12})\simeq1-C_1v_{12}^{1/5}+C_2v_{12}^{2/5}$ \cite{viscoeps}. The
coefficients $C_1$ and $C_2$ depend on material properties and the sizes of gas
particles \cite{REM}. The granular temperature of the viscoelastic gas scales as
$T(t)\sim t^{-5/3}$ \cite{poeschelvisco,Brilveldistrib} implying $D(t)\sim t^{-5/6}$
\cite{seldiffusionbril}. Especially in mixtures of viscoelastic granular particles
a crossover from superdiffusion to a giant diffusion occurs \cite{annaprl,gleb}.

\begin{figure}
\includegraphics[width=8.8cm]{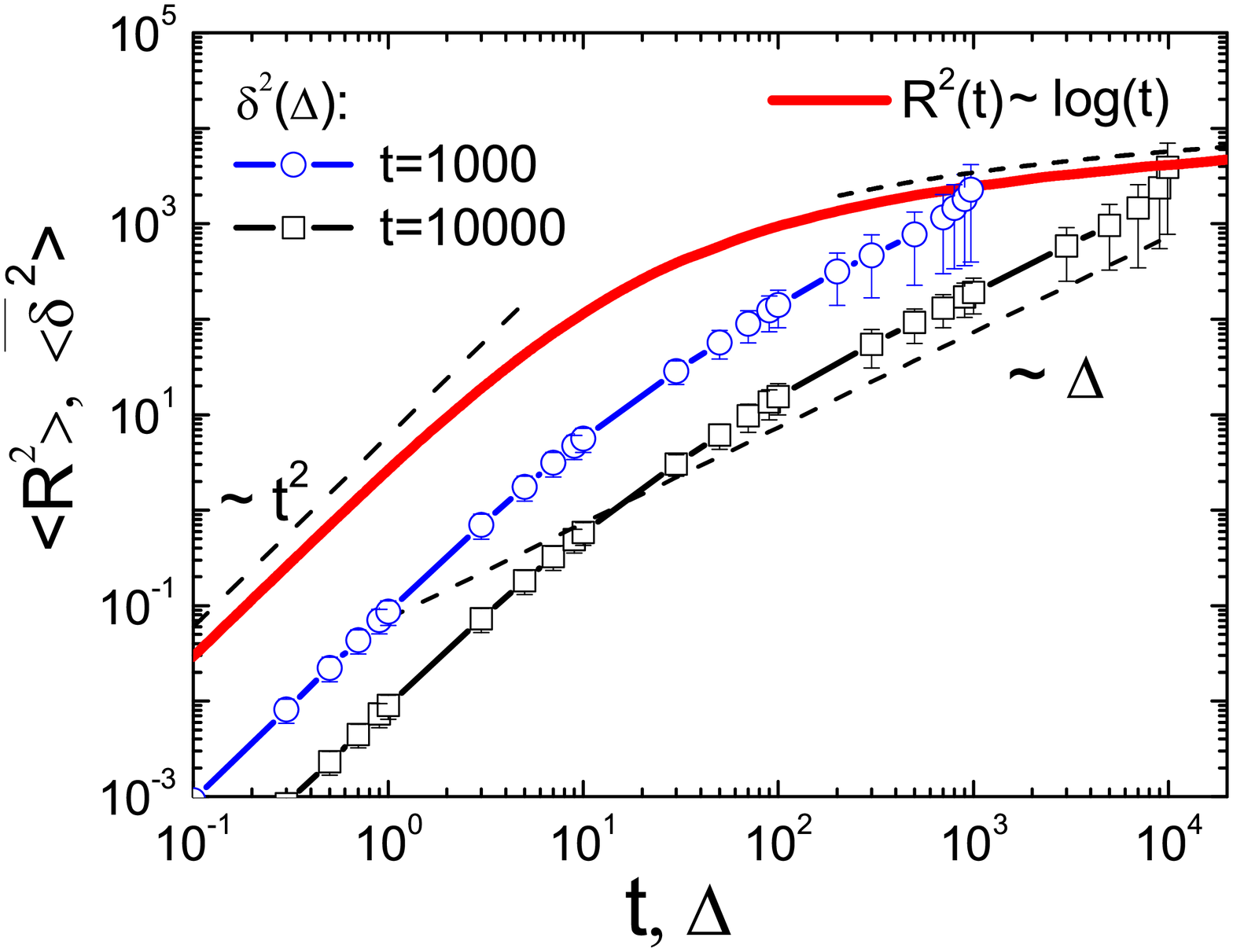}
\includegraphics[width=8.8cm]{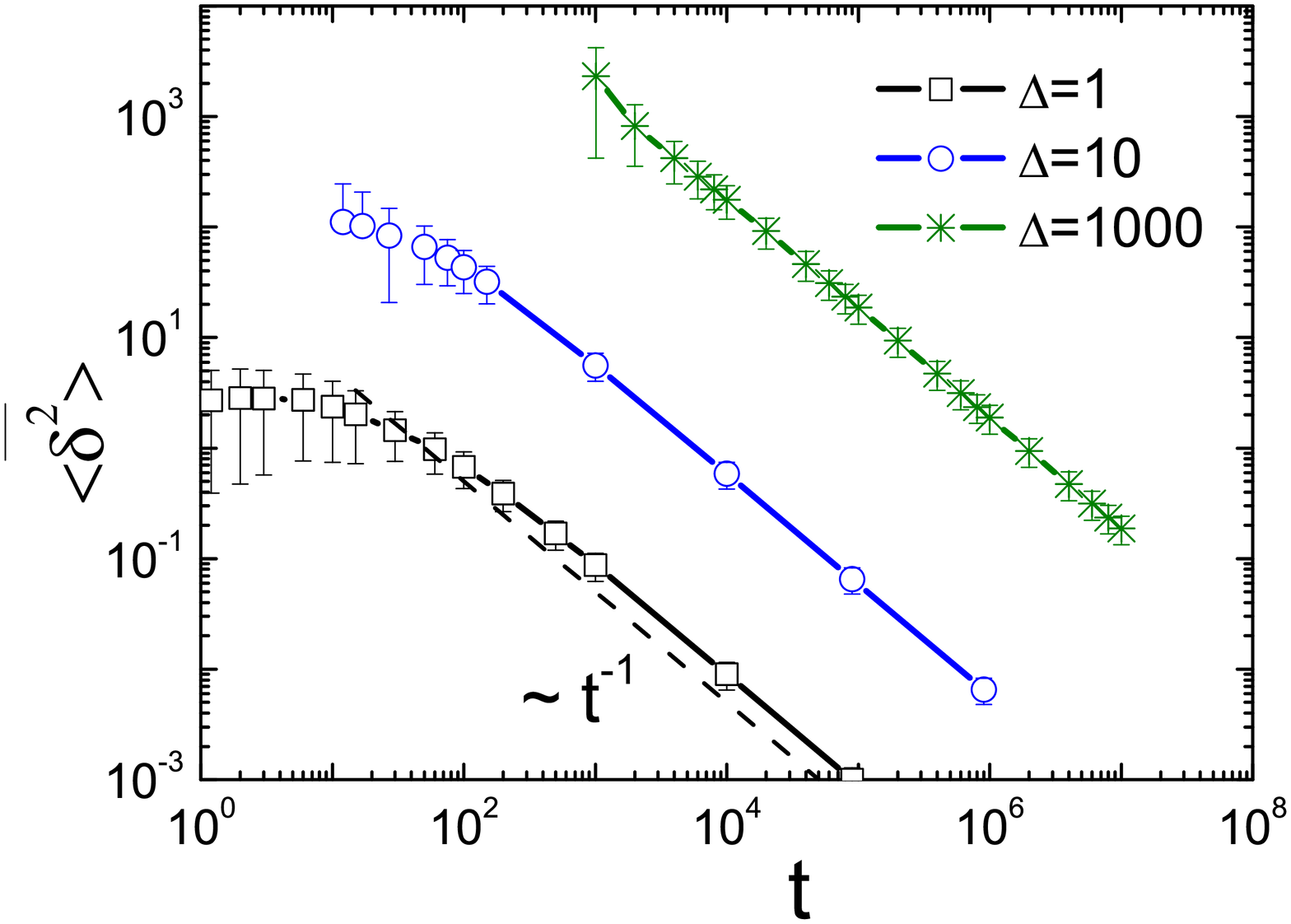}
\caption{Ensemble ($\langle R^2(t)\rangle$) and time averaged ($\langle\overline{
\delta^2(\Delta)}\rangle$) MSDs versus (lag) time (upper graph) and
$\langle\overline{\delta^2(\Delta)}\rangle$ versus length $t$ of the time series
(lower graph), from event-driven MD simulations of a granular gas with
$\varepsilon=0.8$ (symbols). While the ensemble MSD crosses over from ballistic
motion $\left\langle R^2(t)\right\rangle\sim t^2$ for $t\ll\tau_0$ to the
logarithmic law $\left\langle R^2(t)\right\rangle\sim\log(t)$ for $t\gg\tau_0$,
the time averaged MSD starts ballistically and crosses over to
the scaling $\langle\overline{\delta^2(\Delta)}\rangle\sim\Delta/t$ given by
Eq.~(\ref{delta_scaled}).}
\label{GTMSD}
\end{figure}

\begin{figure}
\includegraphics[width=8.8cm]{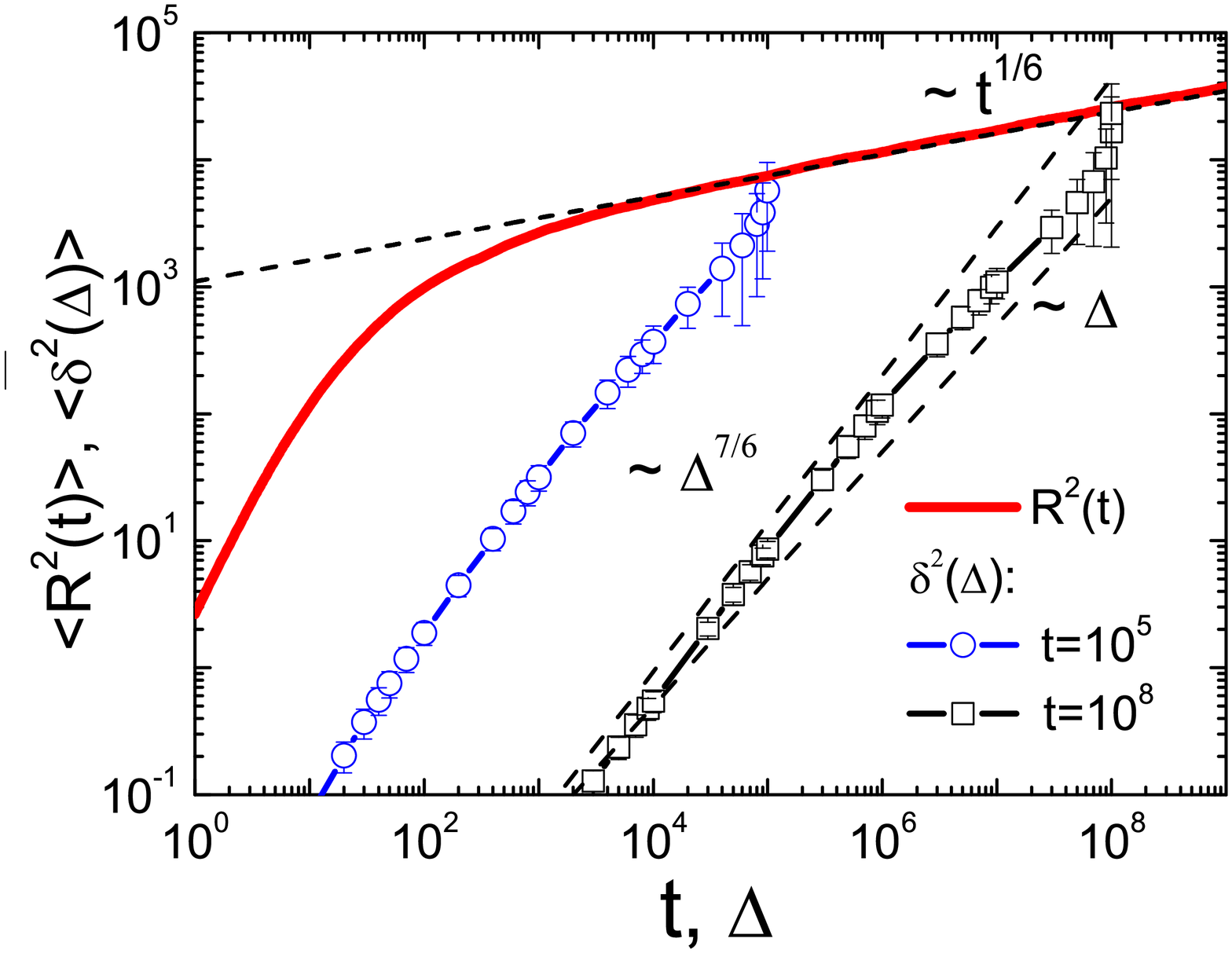}
\includegraphics[width=8.8cm]{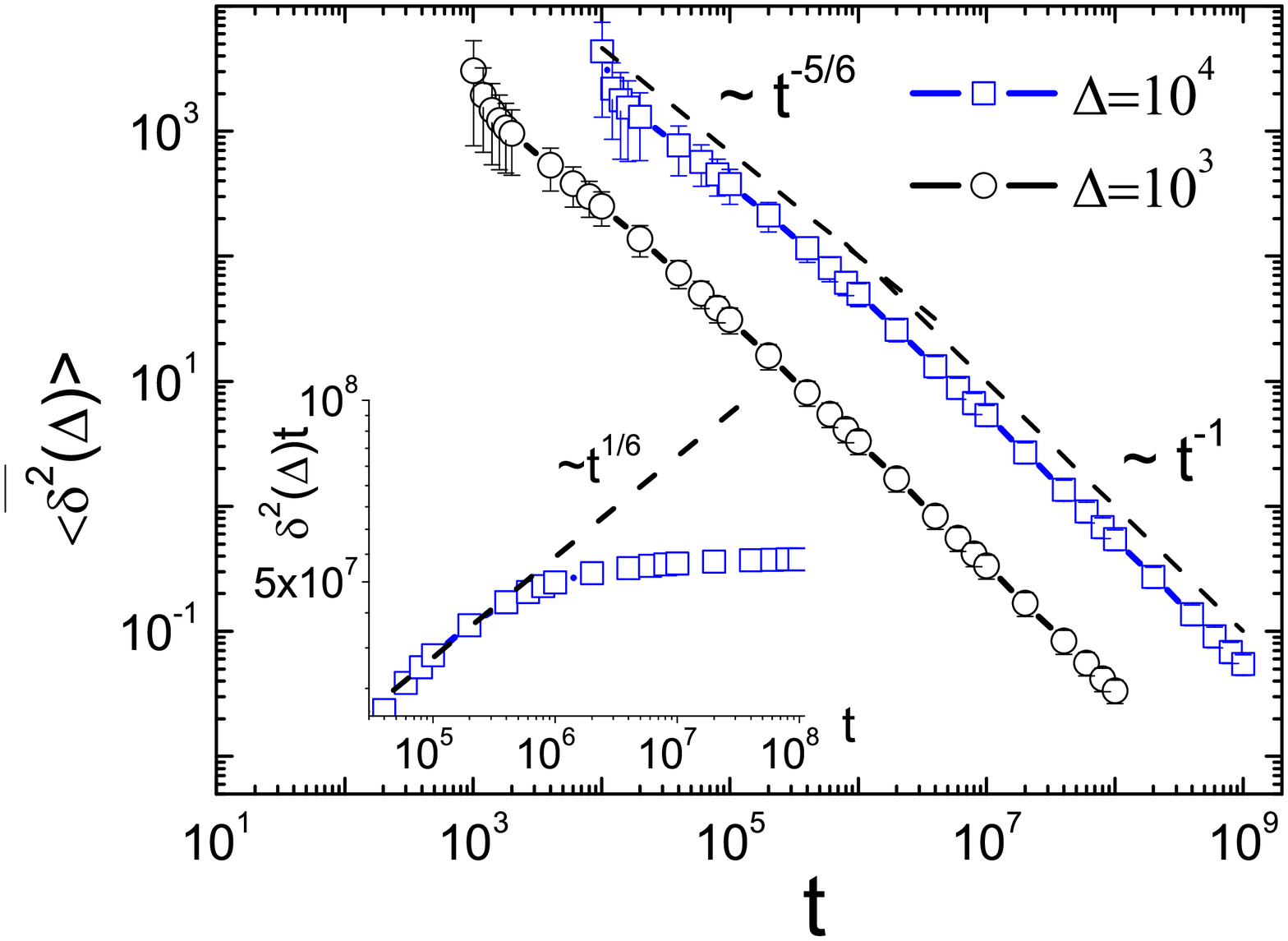}
\caption{MSDs $\langle R^2(t)\rangle$ and $\langle\overline{\delta^2(\Delta)}
\rangle$ as function of (lag) time (top) and $\langle\overline{\delta^2(\Delta)}
\rangle$ versus measurement time $t$ (bottom) from MD simulations (symbols) of a
granular gas with viscoelastic $\varepsilon(v_{12})$. We observe the scaling
$\left\langle R^2(t)\right\rangle\sim t^{1/6}$ for $t\gg\tau_0$. The time
averaged MSD slowly changes between the inidicated slopes (dashed lines).
The continuous change of slope
of $\langle\overline{\delta^2(\Delta)}\rangle$ as function of the length $t$ of
time traces from slope $-5/6$ to $-1$ is seen in the inset of the bottom graph.}
\label{Gvisco}
\end{figure}

\emph{Simulations.} We perform event-driven Molecular Dynamics (MD) simulations
\cite{Compbook} of a gas of hard-sphere granular particles of unit mass and radius,
colliding with constant ($\varepsilon=0.8$, Fig.~\ref{GTMSD}) and viscoelastic
($\varepsilon\left(v_{12}\right)$, Fig.~\ref{Gvisco}) restitution coefficients. The
particles move freely between pairwise collisions, while during the instantaneous
collisions the particle velocities are updated according to certain rules: the
post-collision velocities are given by the velocities before collision and the
restitution coefficient $\varepsilon$. We simulate $N=1000$ particles in a
three-dimensional cubic box with edge length $L=40$ and periodic boundary
conditions. The volume density is $\phi\approx0.065$.

We evaluate the gas dynamics in terms of the standard ensemble MSD $\langle R^2(t)
\rangle$ obtained from averaging over all gas particles at time $t$, as well as
the time averaged MSD
\begin{equation}
\left\langle\overline{\delta^2(\Delta)}\right\rangle=\frac{1}{t-\Delta}\int_0^{
t-\Delta}\left\langle\left[\mathbf{R}(t^{\prime}+\Delta)-\mathbf{R}(t^{\prime})
\right]^2\right\rangle dt'
\label{taMSD}
\end{equation}
for a time series $\mathbf{R}(t)$ of length $t$ as function of the lag time
$\Delta$. Here the angular brackets denote the average $\langle\overline{\delta^2(
\Delta)}\rangle=(1/N)\sum_i^N\overline{\delta^2_i(\Delta)}$ over all $N$ particle
traces. For an ergodic system, such as an ideal gas with unit restitution
coefficient corresponding to normal particle diffusion, ensemble and time averaged
MSDs are equivalent, $\langle R^2(\Delta)\rangle=\langle\overline{\delta^2(\Delta)}
\rangle$ \cite{m1,mreview}. In contrast, several systems characterized by anomalous
diffusion with power-law MSD $\langle R^2(t)\rangle\simeq t^{\alpha}$ ($\alpha\neq
1$) or a corresponding logarithmic growth of the MSD, are nonergodic and display the
disparity $\langle R^2(\Delta)\rangle\neq\langle\overline{\delta^2(\Delta)}
\rangle$ \cite{mreview,glass,m1,sinai,m2,metzlerprl08,schulz}.

Fig.~\ref{GTMSD} shows the results of our simulations of a granular gas with
constant $\varepsilon=0.8$. The ensemble MSD shows initial ballistic particle
motion, $\left\langle R^2(t)\right\rangle\sim t^2$. Eventually, the particles
start to collide and gradually lose kinetic energy. The ensemble MSD of the gas
in this regime follows the logarithmic law $\left\langle R^2(t)\right\rangle\sim
\log(t)$ (red line in Fig.~\ref{GTMSD}, top) \cite{book} (see also Supplementary
Material (SM) \cite{supp}). The time averaged MSD at short lag times $\Delta$
preserves the ballistic law $\langle\overline{\delta^2(\Delta)}\rangle\sim\Delta^2$.
At longer lag times, we observe the linear growth $\langle\overline{\delta^2(\Delta)
}\rangle\sim\Delta$ (black symbols in Fig.~\ref{GTMSD}, top). In addition to this
nonergodic behavior, the time averaged MSD \emph{decreases\/}
with increasing length $t$ of the recorded trajectory, $\langle\overline{\delta^2(
\Delta)}\rangle\sim1/t$. This highly non-stationary behavior is also referred to as
aging, the dependence of the system on its time of evolution \cite{schulz}.
It implies that the system is becoming progressively slower. We observe the
convergence $\lim_{\Delta\to t}\langle\overline{\delta^2(\Delta)}\rangle
\to\left\langle R^2(t)\right\rangle$.

Fig.~\ref{Gvisco} depicts the MD simulations results for a granular gas with
viscoelastic restitution coefficient $\varepsilon(v_{12})$. In this case the
ensemble MSD scales as $\left\langle R^2(t)\right\rangle\sim t^{1/6}$ for $t\gg
\tau_0$. The time averaged MSD does not seem to follow a universal scaling law
but appears to transiently change from the power-law $\langle\overline{\delta^2}
\rangle\sim\Delta^{7/6}$ at intermediate lag times to $\langle
\overline{\delta^2}\rangle\sim\Delta$ at longer $\Delta$, see the bounds
derived in SM \cite{supp}. As function of the length $t$ of particle traces, we
observe the crossover from $\langle\overline{\delta^2}\rangle\sim t^{-5/6}$ to
$\langle\overline{\delta^2}\rangle\sim1/t$.

\emph{Granular gas with constant $\varepsilon$.} Let us explore this behavior in
more detail. The dynamics of a granular gas can be mapped to that of a molecular
gas by a rescaling of time from $t$ to $\tau$ as $d\tau=\sqrt{T(t)/T(0)}dt$
\cite{scale}. Using Haff's law (\ref{haff}), it follows that $\tau=\tau_0\log\left(
1+t/\tau_0\right)$. As function of $\tau$ the granular temperature remains constant,
and the velocity-velocity correlation function for each spatial component decays
exponentially \cite{scale,book},
\begin{equation}
\langle v(\tau_1)v(\tau_2)\rangle=T_0/m\times\exp(-|\tau_2-\tau_1|/\tau_v(0)).
\label{velgran}
\end{equation}
The MSD then has the exponential approach $\langle R^2(\tau)\rangle=6D_0\tau-6D_0
\tau_v(0)(1-\exp[-\tau/\tau_v(0)])$ to the Brownian law $\langle R^2(\tau)\rangle
=6D_0\tau$ \cite{scale,book}. In real time $t$ we find ($t_2\ge t_1$)
\begin{equation}
\langle v(t_1)v(t_2)\rangle=\frac{T_0}{m}\left(1+t_1/\tau_0\right)^{\beta-1}
\left(1+t_2/\tau_0\right)^{-\beta-1}
\label{vcorgran}
\end{equation}
for the velocity correlator ($\beta=\tau_0/\tau_v(0)$). The MSD is
\begin{eqnarray}
\nonumber
\left\langle R^2(t)\right\rangle&=&6D_0[\tau_0\log\left(1+t/\tau_0\right)\\
&&+\tau_v(0)[(1+t/\tau_0)^{-\beta}-1]].
\label{R2granbal}
\end{eqnarray}
At short times the particles move ballistically, $\left\langle R^2(t)\right\rangle
\sim 3D_0t^2/\tau_v(0)$, crossing over to the logarithmic growth $\left\langle R^2
(t)\right\rangle\sim6D_0\tau_0\log\left(t/\tau_0\right)$ as seen in Fig.~\ref{GTMSD},
top.

Based on the velocity autocorrelation (\ref{vcorgran}) we now obtain the time
averaged MSD (see details in SM \cite{supp})
\begin{equation}
\left\langle\overline{\delta^2(\Delta)}\right\rangle\simeq 6D_0\tau_0\Delta/t
\label{delta_scaled}
\end{equation}
valid in the range $\tau_0\ll\Delta\ll t$, where $\tau_0$ is the characteristic
decay time of the temperature law (\ref{haff}). This result indeed explains
the observed behavior of Fig.~\ref{GTMSD}: the time averaged MSD scales linearly
with the lag time and inverse-proportionally with the length $t$ of the time
traces. Comparison of Eqs.~(\ref{R2granbal}) and (\ref{delta_scaled}) demonstrates
the nonergodicity and aging properties of the gas particles.

\textit{Viscoelastic granular gas.} For a velocity-dependent $\varepsilon(v_{12})$
the temperature decays like $T(t)\simeq T_0\left(t/\tau_0\right)^{-5/3}$, and the
time transformation reads $\tau=6\tau_0^{5/6}t^{1/6}$. The MSD in this case exhibits
the long time scaling $\left\langle R^2(t)\right\rangle\sim36D_0\tau_0^{5/6}t^{1/6}$
seen in Fig.~\ref{Gvisco}, top. For the time averaged MSD we analytically obtain the
bounds $\langle\overline{\delta^2(\Delta)}\rangle\sim\Delta^{7/6}/t$ and $\langle
\overline{\delta^2(\Delta)}\rangle\sim\Delta/t^{5/6}$, compare the details
in the SM \cite{supp}. These bounds are given by the dashed lines in
Fig.~\ref{Gvisco}, top. Concurrent to this change of slopes as function of the lag
time, Fig.~\ref{Gvisco}, bottom, shows the change of slope of $\langle\overline{
\delta^2(\Delta)}\rangle$ as function of the trajectory length $t$ from the slope
$-5/6$ to $-1$ at a fixed lag time.

\textit{Scaled Brownian motion.} For unit restitution coefficient individual
gas particles at long times perform Brownian motion at a fixed temperature defined
by the initial velocity distribution. For the dissipative granular gases considered
herein, the temperature scales like $T(t)\simeq1/t^2$ and $\simeq1/t^{5/3}$,
respectively. Single-particle stochastic processes with power-law time-varying
temperature or, equivalently, time dependent diffusivity $D(t)$, are well known.
Such SBM is described in terms of the overdamped Langevin
equation with diffusivity $D(t)\sim t^{\alpha-1}$ for $0<\alpha<2$ \cite{sbm,sbm1}.
SBM is highly non-stationary and known to be nonergodic and aging
\cite{sbm1,sbm_age,mreview}.

To study whether SBM provides an effective single-particle description of the
diffusion in dissipative granular gases we extend SBM to the underdamped case,
\begin{equation}
\label{sbm}
d\mathbf{v}/dt+\mathbf{v}/[\tau_v(t)]=\sqrt{2D(t)}/\tau_v(t)\times\boldsymbol{
\xi}(t)
\end{equation}
driven by white Gaussian noise $\boldsymbol{\xi}(t)$ with correlation function
$\langle\xi_i(t_1)\xi_j(t_2)\rangle=\delta_{i,j}\delta(t_1-t_2)$ for the components.

A granular gas with constant $\varepsilon<1$ corresponds to the limiting case
$\alpha=0$, giving rise to the logarithmic form $\langle R^2(t)\rangle\simeq D_0
\log(t)$ and the velocity correlation \cite{ultraslow_sbm}
\begin{equation}
\left\langle v(t_1)v(t_2)\right\rangle=\frac{T(0)\tau_0}{m\tau_v(0)(\beta-1)}
\frac{(1+t_1/\tau_0)^{\beta-2}}{(1+t_2/\tau_0)^{\beta}}.
\end{equation}
This result for the ultraslow SBM formally coincides with the velocity correlation
function (\ref{vcorgran}) for granular gases in the limit $\beta\gg1$, in which the
velocity correlation time $\tau_v(0)$ is much smaller than the characteristic decay
time $\tau_0$ of the granular temperature. This is achieved for sufficiently small
dissipation in the system ($\varepsilon\lessapprox1$).

A more careful analysis shows that there exists a difference of the SBM model to
the full result for the granular gas due to
ballistic correlations \cite{supp}: The restitution coefficient affects only the
normal component of the velocity of the colliding particles, while the tangential
components remain unchanged. Therefore, in a granular gas with constant restitution
coefficient the trajectories of particles become more and more aligned, effecting
long-termed correlations. In the case of viscoelastic particles the restitution
coefficient tends to unity when the relative velocities of colliding particles
decrease with time. Therefore, at long times, the trajectories become more chaotic,
and the ballistic correlations disappear. In this viscoelastic case ballistic
correlations play a significant role only for intermediate lag times, while they
become suppressed at longer lag times, effecting a gradual crossover between the
bounding behaviors $\left\langle\overline{\delta^2(\Delta)}\right\rangle\simeq
\Delta^{7/6}/t$ and $\simeq\Delta/t^{5/6}$ (Fig.~\ref{Gvisco}, top).
In contrast, for a constant $\varepsilon$ the ballistic correlations are relevant
during the entire evolution of the gas, canceling the leading term of the SBM
result for the time averaged MSD \cite{supp}.

\textit{Conclusions.} The occurrence of nonergodicity in the form of the disparity
between (long) time and ensemble averages of physical observables and aging, is
not surprising in strongly disordered systems described by the prominent class of
continuous time random walk models involving divergent time scales of the dynamics
\cite{mreview,m1,glass,metzlerprl08}. Examples include diffusive motion in amorphous
semiconductors, structured disordered environments, or living biological cells
\cite{mreview}.

In contrast to such complex systems, we here demonstrated how nonergodicity
arises in force-free granular gases. This may seem surprising for such simple
mechanistic systems. However, the physical reason for the nonergodicity is due to
the strong non-stationarity brought about by the continuous decay of the gas
temperature. For a constant restitution coefficient the MSD $\langle R^2(t)\rangle$
grows logarithmically, while the time averaged MSD $\langle\overline{\delta^2
(\Delta)}\rangle$ is linear in the lag time and decays inverse proportionally with
the length of the analyzed time traces (aging).
We derived the observed nonergodicity and the aging behavior of granular gases
from the velocity autocorrelation functions. We note that aging in the
homogeneous cooling state of granular gases was reported previously \cite{age},
however, it was not put in context with the diffusive dynamics of gas particles.

The decaying
temperature of the dissipative force-free granular gas corresponds to an increase
of the time span between successive collisions of gas particles, a feature directly
built into the SBM model \cite{sbm1}. As we showed here, SBM and its ultraslow
extension with the logarithmic growth of the MSD indeed captures certain features
of the observed motion and may serve as an effective single-particle model for the
granular gas, which is particularly useful when more complex situations are
considered, such as the presence of external force fields.

Granular gases represent a fundamental physical system in statistical mechanics,
extending the ideal gas model to include dissipation on particle collisions.
Granular gases are a reference model in granular matter physics \cite{mehta}
with applications ranging from interstellar clouds or planetary rings to
technologies in food and construction industries \cite{book,mehta}. Our results
shed new light on the physics of granular gases with respect to their violation
of ergodicity in the Boltzmann sense. Moreover, we obtained the detailed influence
of ballistic correlations in the granular gas dynamics. Both results are important
for a better understanding of dissipation in free gases as well as the analysis
of experimental observations and MD studies of granular gases.

\begin{acknowledgments}
We thank F.  Spahn, I. M. Sokolov and N. V. Brillian\-tov for discussions.
Simulations were run at Moscow State University's Chebyshev supercomputer.
We acknowledge support from EU IRSES grant DCP-PhysBio N269139, IMU Berlin
Einstein Foundation, DFG grant CH707/5-1, and Academy of Finland
(FiDiPro scheme to RM).
\end{acknowledgments}

\clearpage

\onecolumngrid

\renewcommand{\theequation}{S\arabic{equation}}
\renewcommand{\thefigure}{S.\arabic{figure}}

\begin{center}

\textbf{\large Supplementary Material: Nonergodic dynamics of force-free granular
gases}\\[0.4cm]

Anna Bodrova$^{1,2}$, Aleksei V. Chechkin$^{1,3}$, Andrey G. Cherstvy$^1$, and
Ralf Metzler$^{1,4}$\\[0.12cm]

\textit{$^1$Institute of Physics and Astronomy, University of Potsdam,
D-14476 Potsdam-Golm, Germany\\
$^2$Faculty of Physics, M. V. Lomonosov Moscow State University, Moscow 119991,
Russia\\
$^3$Akhiezer Institute for Theoretical Physics, Kharkov Institute of Physics
and Technology, Kharkov 61108, Ukraine\\
$^4$Department of Physics, Tampere University of Technology, 33101 Tampere,
Finland}\\[0.4cm]

\noindent
\begin{minipage}{14.4cm}
In this Supplemental Material (SM) we present the details of the derivation of the
results presented in the main manuscript as well as an additional Figure
comparing the results for a viscoelastic granular gas with the numerical
evaluation of the time averaged mean squared displacement.
\end{minipage}

\end{center}

\section{Granular gas: constant restitution coefficient}

The time-averaged mean-squared displacement (MSD) for the granular gas with constant
restitution coefficient, see Eq.~(3) in the main text, may be presented in the
following form
\begin{equation}
\left\langle\overline{\delta^2(\Delta)}\right\rangle=\frac{1}{t-\Delta}\int_0^{t-\Delta}dt^{\prime}\left[\left\langle R^2(t^{\prime}+\Delta)\right\rangle - \left\langle R^2(t^{\prime})\right\rangle - 2A\left(t^{\prime},\Delta\right)\right],
\label{StaMSD}
\end{equation} 
where the MSD $\left\langle R^2(t)\right\rangle$ is defined according to Eq.~(6) of the main text and 
\begin{equation}\label{AAA}
A\left(t,\Delta\right)=3\int_0^t dt_1 \int_t^{t+\Delta} dt_2 \langle v_x(t_1)v_x(t_2)\rangle =
\frac{3T_0\tau_v^2(0)}{m}\left[1-\left(1+\frac{t}{\tau_0}\right)^{-\beta}-\left(1+\frac{\Delta}{\tau_0+t}\right)^{-\beta}+\left(1+\frac{t+\Delta}{\tau_0}\right)^{-\beta}\right]\,.
\end{equation}
This term accounts for position correlations at different time moments $t$
and $t+\Delta$. It vanishes for underdamped scaled Brownian motion (SBM) governed by
the Langevin equation as well as for ultraslow SBM (Refs.~[34,36] of the
main text). In the present consideration, this term is non-zero because
there are long-term ballistic correlations in a granular gas with a constant
restitution coefficient $\epsilon$. It arises due to the fact that the normal
component of the relative velocity of colliding particles decreases, while the
tangential one remains unchanged in the course of collisions, see Ref. [2]
of the main text. Introducing Eq.~(\ref{AAA}) and Eq.~(6) from the main
text into Eq.~(\ref{StaMSD}), we get the time-averaged MSD in the form
\begin{eqnarray}
\label{delta01}
\left\langle\overline{\delta^2(\Delta)}\right\rangle=\left\langle\overline{\delta_0^2(\Delta)}\right\rangle+\Xi(\Delta)\,.
\end{eqnarray}
The first part is equal to the time-averaged MSD obtained for the overdamped SBM
\begin{eqnarray}
\nonumber
\left\langle\overline{\delta_0^2(\Delta)}\right\rangle&=&\frac{6D_0\tau_0}{t-\Delta}
\int_0^{t-\Delta}dt^{\prime}\log\left(\frac{\tau_0+t^{\prime}+\Delta}{t^{\prime}+
\tau_0}\right)\\
&=&\frac{6D_0\tau_0}{t-\Delta}\left[\left(t+\tau_0\right)\log\left(t+\tau_0\right)-
\left(\Delta+\tau_0\right)\log\left(\Delta+\tau_0\right)-\left(t-\Delta+\tau_0\right)
\log\left(t-\Delta+\tau_0\right)+\tau_0\log\tau_0\right].
\label{delta00}
\end{eqnarray}
For $\tau_0\ll\Delta\ll t$
\begin{equation}
\left\langle\overline{\delta_0^2(\Delta)}\right\rangle\simeq\frac{6D_0\tau_0\Delta}{t}\left[\log\left(\frac{t}{\Delta}\right)+1\right].
\end{equation}
The second part has the form
\begin{eqnarray}\nonumber
\Xi(\Delta)=\frac{6D_0\tau_v(0)}{t-\Delta}\int_{0}^{t-\Delta}dt^{\prime}\left[\left(1+\frac{\Delta}{t^{\prime}+\tau_0}\right)^{-\beta}-1\right]<0,
\end{eqnarray}
where $\beta=\tau_0/\tau_v(0)$, see the main text. Introducing the new variable $y=\Delta/t^{\prime}$ we get in the limit $\tau_0\ll\Delta$ that 
\begin{equation}
\Xi\left(\Delta\right)\simeq-6D_0\tau_v(0)\left(1-\frac{\Delta}{t-\Delta}I\left(t,\Delta\right)\right), \,\,\textrm{where} \,\,\,I\left(t,\Delta\right)=\int_{\frac{\Delta}{t-\Delta}}^{\infty}\frac{dy}{y^2\left(1+y\right)^{\beta}}\,.\end{equation}
This integral can be taken by parts
\begin{equation}
I\left(t,\Delta\right)=\frac{t-\Delta}{\Delta}\left(1-\frac{\Delta}{t}\right)^{\beta}+\beta\log\left(\frac{\Delta}{t-\Delta}\right)\left(1-\frac{\Delta}{t}\right)^{\beta+1}+\beta C\left(\beta\right),
\end{equation}
where
\begin{equation}
C\left(\beta\right)=-\left(\beta+1\right)\int_{\frac{\Delta}{t-\Delta}}^{\infty}dy\frac{\log y}{\left(1+y\right)^{\beta+2}}\simeq\gamma+\frac{1}{\beta}+\psi\left(\beta\right)\,,
\end{equation}
 $\gamma=0.5772...$ is the Euler's constant, and $\psi\left(z\right)=\frac{d\log\Gamma(z)}{dz}$ is the so-called digamma function.
Finally, we get in the limit of $t\gg\Delta$ that
\begin{eqnarray}
\left\langle\overline{{\delta}^2(\Delta)}\right\rangle \simeq 6D_0\tau_0 C\left(\beta\right)\frac{\Delta}{t}.
\label{Sdelta_scaled}
\end{eqnarray}
This confirms the linear scaling as function of the lag time of the time averaged
MSD and thus the nonergodic nature of the granular gas.

\section{Granular gas: velocity-dependent restitution coefficient}

Similarly, for the viscoelastic granular gas with $\epsilon=\epsilon(v_{12})$
the time-averaged MSD may be
presented as the sum of two parts, see Eq.~(\ref{delta01}). The first
term corresponds to the time-averaged MSD of the SBM process, which can be
described by the overdamped Langevin equation, namely
\begin{eqnarray}
\left\langle\overline{\delta_0^2(\Delta)}\right\rangle=\frac{36D_0\tau_0^{5/6}}{t-\Delta}\int_0^{t-\Delta}dt^{\prime}\left[\left(t^{\prime}+\Delta\right)^{1/6}-t^{\prime 1/6}\right]=\frac{216D_0\tau_0^{5/6}}{7\left(t-\Delta\right)}\left[t^{7/6}-\Delta^{7/6}-\left(t-\Delta\right)^{7/6}\right]\,.
\label{eq-tamsd-sbm}\end{eqnarray}
The second term, accounting for ballistic correlations, becomes
\begin{eqnarray}
\Xi(\Delta)=\frac{6D_0\tau_v(0)}{t-\Delta}\int_0^{t-\Delta}dt^{\prime}\left[\exp\left(-\frac{6\tau_0^{5/6}}{\tau_v(0)}\left[\left(t^{\prime}+\Delta\right)^{1/6}-t^{\prime 1/6}\right]\right)-1\right]\,.
\label{eq-addition-theta}\end{eqnarray}
This integral can be presented as a sum of three parts
\begin{eqnarray}
\label{Id1}
\int_0^{t-\Delta}dt^{\prime}\left[\exp\left(-\frac{6\tau_0^{5/6}}{\tau_v(0)}\left[\left(t^{\prime}+\Delta\right)^{1/6}-t^{\prime 1/6}\right]\right)-1\right]=\int_0^{k_1\Delta}[\dots]dt^{\prime}+\int_{k_1\Delta}^{k_2\Delta}[\dots]dt^{\prime}+\int_{k_2\Delta}^{t-\Delta}[\dots]dt^{\prime}\,.
\end{eqnarray}
We choose the coefficients $k_{1,2}$ in the following ranges  \begin{equation}1\ll k_1\ll \tau_0\Delta^{1/5}/\tau_v^{6/5}(0) \textrm{~~and~~}  \tau_0\Delta^{1/5}/\tau_v^{6/5}(0)\ll k_2 \ll t/\Delta.\label{eq-range}\end{equation} This enables us to evaluate the first integral in Eq.~(\ref{Id1}) as follows
\begin{equation}
\int_0^{k_1\Delta}dt^{\prime}\left[\exp\left(-\frac{6\tau_0^{5/6}}{\tau_v(0)}\left[\left(t^{\prime}+\Delta\right)^{1/6}-t^{\prime 1/6}\right]\right)-1\right]\sim-k_1\Delta\,.
\label{eq-first-term}\end{equation}
The third term in Eq. (\ref{Id1})
can be evaluated as

\begin{eqnarray}
\nonumber
\int_{k_2\Delta}^{t-\Delta}dt^{\prime}\left[\exp\left(-\frac{6\tau_0^{5/6}}{\tau_v(
0)}\left[\left(t^{\prime}+\Delta\right)^{1/6}-t^{\prime 1/6}\right]\right)-1\right]\\
&&\hspace*{-9cm}
\simeq\int_{k_2\Delta}^{t-\Delta}dt^{\prime}\left(-\frac{6\tau_0^{5/6}}{\tau_v(0)}
\left[\left(t^{\prime}+\Delta\right)^{1/6}-t^{\prime 1/6}\right]\right)
=\frac{36\tau_0^{5/6}}{7\tau_v(0)}\left[-t^{7/6}+\left(t-\Delta\right)^{7/6}+\left(
\left(k_2+1\right)^{7/6}-k_2^{7/6}\right)\Delta^{7/6}\right].
\label{II2}
\end{eqnarray}
Under the assumptions (\ref{eq-range}) the contribution of the terms (\ref{eq-first-term}) can be neglected. Finally, assuming that the second term in Eq. (\ref{Id1})
is small enough compared to Eqs. (\ref{II2}) for the range  of parameters $k_{1,2}$ chosen, we get to leading order
\begin{equation}
\left\langle\overline{{\delta}^2(\Delta)}\right\rangle \simeq 36k_2^{1/6}D_0\tau_0^{5/6}\frac{\Delta^{7/6}}{t}\,.
\label{delta_visco}
\end{equation}

For larger values of the lag time $\Delta$, in the range $\tau_v(0)
t^{5/6}/\tau_0^{5/6}\ll\Delta\ll t$ that is opposite to the second condition in Eq. (\ref{eq-range})  we have the upper estimate for the correction $\Xi(\Delta)$ to the time averaged MSD of  the SBM process, namely 
\begin{equation}
\left|\Xi(\Delta)\right|\le 6D_0\tau_v(0)\ll \left\langle\overline{\delta_0^2(\Delta)}\right\rangle \,.
\end{equation}Then we get in the limit $\Delta\ll t$ that 
\begin{equation}
\left\langle\overline{\delta^2(\Delta)}\right\rangle\sim  \left\langle\overline{\delta_0^2(\Delta)}\right\rangle \,\simeq\frac{D_0\tau_0^{5/6} 
\Delta}{t^{5/6}}\,.
\label{eq-sbm-scaling}\end{equation}

\begin{figure}
\includegraphics[width=12cm]{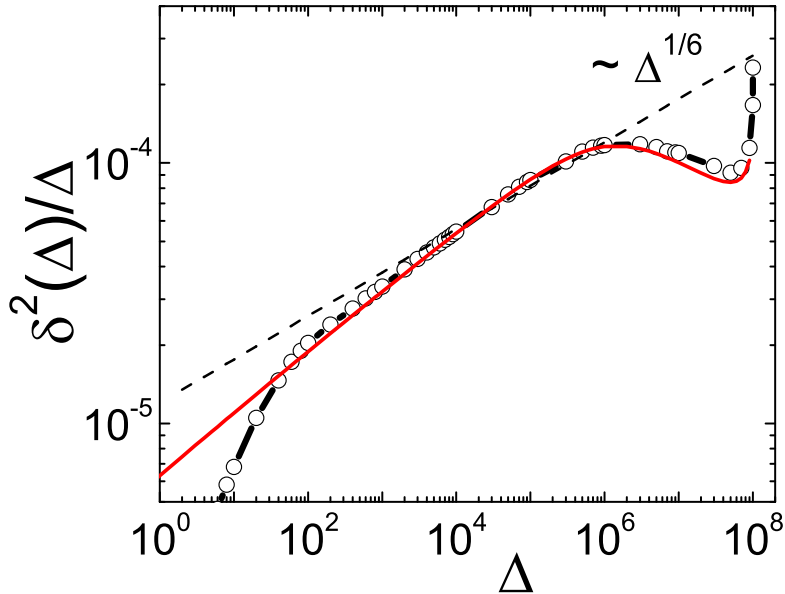}
\caption{Time averaged MSD $\langle\overline{\delta^2(\Delta)}\rangle$ divided
by lag time $\Delta$ as function of $\Delta$ from MD simulations (symbols)
of a granular gas with velocity-dependent restitution coefficient. The lines connecting the symbols guide the eye. Red line corresponds
to numerical calculation of $\langle\overline{\delta^2(\Delta)}\rangle$ in Eq. (\ref{delta01}) for
$\tau_0=25$, $\tau_v=2$, $D_0=2$. These values ensure the closest agreement and are consistent with the parameters of the granular gas as used in MD simulations. Dashed line shows the asymptotic
$\langle\overline{\delta^2(\Delta)}\rangle/\Delta\sim\Delta^{1/6}$ behavior according to Eq. (\ref{eq-sbm-scaling}). }
\label{Gdeltainset}
\end{figure}

In addition to these analytical estimates, we computed numerically the full expression for the time averaged MSD $\left\langle\overline{\delta^2(\Delta)}\right\rangle= \left\langle\overline{\delta_0^2(\Delta)}\right\rangle+\Xi(\Delta)$. It agrees well with our MD simulation data, compare the curves in 
Fig.~\ref{Gdeltainset} where we explicitly plot $\left\langle\overline{\delta^2}\right\rangle/\Delta$. It shows that in the range of parameters $\tau_0,~\tau_v$ and $D_0$ consistent with the MD\ simulations presented in the main text, the transient scaling behavior  $\left\langle\overline{\delta^2(\Delta)}\right\rangle\sim \Delta^{7/6}$ is realized in a limited range of $\Delta$. Note also that in this range the linear SBM scaling for $\left\langle\overline{\delta^2(\Delta)}\right\rangle$ as prescribed by Eq. (\ref{eq-sbm-scaling}) is no longer valid. The reason is that for large values of $\Delta,$  when $\Delta\to t$ for any length of the trajectory, in Eq. (\ref{eq-tamsd-sbm}) the evolution of the time averaged MSD with the lag time $\Delta$ becomes inherently nonlinear.

\begin{thebibliography}{29}
\bibitem{REMM} In this definition of the restitution coefficient we consider the
normal component of the relative particle velocity. Rough particles are also
characterized by a tangential restitution coefficient \cite{book}.
\bibitem{REM} The constants are given by $C_1=1.15A\kappa^{2/5}$ and $C_2=0.798
A^2\kappa^{4/5}$, where $A$ quantifies the specific viscous materials properties
of the gas granules, and $\kappa=\left(\frac{3}{2}\right)^{3/2}Y\sigma^{1/2}/[m(
1-\nu^2)]$ is the elastic constant. The latter involves the Young's modulus $Y$,
the Poisson ratio $\nu$, and the diameter $\sigma$ of the particles \cite{book}.
\end{thebibliography}
\end{document}